\documentclass[12pt]{article}
\usepackage[dvips]{graphicx}
\usepackage{rangecite}
\usepackage{amssymb}
\def\time-stamp{Time-stamp: "2000-11-03 16:03:42 yoko"}

\usepackage{calc}
\newlength{\pagewidthhalf}
\newfont{\bg}{cmr10 scaled\magstep4}

\newcommand{\bigzerou}{\smash{\lower1.7ex\hbox{\bg 0}}}
\begin{document}
\begin{titlepage}
{\begin{center}
{\Large\bf 
Matter Enhancement of T Violation \\
\vspace*{0.3em}
 in Neutrino Oscillation
}
\end{center}}
\renewcommand{\thefootnote}{\alph{footnote}}
\vspace*{1em}
\begin{center}
{\large\sc 
H.~Yokomakura\footnote[1]{\makebox[1.cm]{Email:}
yoko@eken.phys.nagoya-u.ac.jp}
 and 
K.~Kimura\footnote[2]{\makebox[1.cm]{Email:}
kimukei@eken.phys.nagoya-u.ac.jp}
\\
 }
\vspace*{1em}
{\it Department of Physics, Nagoya University \\
Nagoya 464-8602, Japan}

\vspace*{2em}
{\large\sc 
A.~Takamura\footnote[3]{\makebox[1.cm]{Email:}
takamura@eken.phys.nagoya-u.ac.jp}}
\\
\vspace*{1em}
{\it Department of Mathematics, Toyota National College of Technology \\
 Eisei-cho 2-1, Toyota-shi, 471-8525, Japan}
\end{center}

\thispagestyle{empty}

\section*{ }
        \begin{center}
       -- Abstract --
        \end{center}

We study the matter enhancement of T violation
 in neutrino oscillation with three generations. 
 The magnitude of T violation is proportional to
  Jarlskog factor $J$. 
Recently, 
the elegant relation, 
$
(\Delta_m)_{12}
(\Delta_m)_{23}
(\Delta_m)_{31}
J_{m}
=
\Delta_{12}
\Delta_{23}
\Delta_{31}
J
$, 
was derived, 
where 
 $\Delta_{ij}=\Delta m^2_{ij}/(2E)$
  and
 subscript $m$ implies the quantities in matter. 
Using this relation, 
we reconsider how $J_m$ changes 
 as a function of the matter potential $a$ 
under the approximation 
 $|\Delta m^2_{12}| \ll |\Delta m^2_{13}|$. 
We show that  
the number of maxima for $J_{m}$ 
depends on the magnitude of $\sin^2 2 \theta_{13}$ 
and 
 there are two maxima 
 considering the constraint on $\sin^2 2\theta_{13}$ from 
 the CHOOZ experiment. 
 One maximum of $J_{m}$ 
 at $a = O(\Delta_{12})$ 
 is given by 
 $J/\sin2\theta_{12}$, 
 which leads to the large enhancement of $J_m$ 
in the case of the SMA MSW solution. 
The other maximum at $a = O(\Delta_{13})$ 
 is 
 $|\Delta_{12}/\Delta_{13}|J/\sin2\theta_{13}$, and 
the enhancement is possible, 
if $\sin 2 \theta_{13}$ is small enough. 
These maximal values are consistent
 with the results obtained by other methods. 

\end{titlepage}
\section{Introduction}
\pagenumbering{arabic} 

\hspace*{\parindent}
Solar neutrino experiments 
have been observing
 a $\nu_e$ deficit for a long time
\cite{solar}
and the ratio of $\nu_\mu/\nu_e$ in atmospheric neutrino
has implied 
 a $\nu_\mu$
 deficit
\cite{atmo}, 
 which are explained by $\nu_e$-$\nu_\mu$ oscillation
 and $\nu_\mu$-$\nu_\tau$ oscillation, respectively. 
These experiments
 provide strong evidence that there exist masses and 
 mixings in the lepton sector with three generations
\cite{MNS}. 

Long baseline experiments \cite{exp} and neutrino factories \cite{n-f-ex-review}
 are operated, or planned, 
 in order to obtain 
 more convincing evidence for neutrino
oscillation. 
Furthermore it could also be possible 
 to observe CP and T violations. 

As 
 the neutrinos
pass through the earth
 in these experiments, 
matter effects must 
 be considered. 
It has been studied
 in the context of 
 long baseline experiments \cite{CP-T},
 and 
 in the context of a neutrino factory \cite{n-f-CP-matter}. 
T violation is different from CP violation in matter
 and it is pointed out that it is easy to calculate 
 T violation compared with CP violation
 for neutrino oscillation in matter \cite{T}. 
 The T violating part in matter, 
$
\Delta P_T
=P(\nu_\alpha \to \nu_\beta)-P(\nu_\beta \to \nu_\alpha)$, 
$(\alpha, \beta=e, \mu, \tau)$
is proportional to Jarlskog factor $J_{m}$ \cite{Jarlskog} of the lepton sector,
unlike the CP violating part. 
The dependence of $J_m$ on the matter potential
$a=\sqrt2 G_F N_e$ 
 is investigated 
 in other works \cite{Xing,Yasuda}. 

Recently, Harrison and Scott \cite{harrison-scott}
 derived the relation  
\begin{eqnarray}
  \label{eq:jmdm-jd}
  (\Delta_m)_{12} (\Delta_m)_{23} (\Delta_m)_{31} J_{m}
 = \Delta_{12} \Delta_{23} \Delta_{31} J, 
\end{eqnarray}
where $\Delta_{ij}=\Delta m^2_{ij}/(2E)$ and 
 the quantities with the subscript $m$ are those in matter.   
  The inverse of $J_{m}$ is the square root of a quartic function
 of $a$. 
This means that 
$J_{m}$ has either one or two local maxima
as a function of $a$. 

In this letter, 
 we present both the exact and approximate 
 form of $J_{m}$ as a function of $a$ using the above relation.
It is shown that the number of
resonant maxima of $J_{m}$ 
 depends on the magnitude of $\sin^2 2 \theta_{13}$. 
Taking account of 
 the constraint on $\sin^2 2 \theta_{13}$ from 
 the CHOOZ experiment \cite{CHOOZ}, 
 we show that there exist two maxima. 
We also estimate the maximal values of $J_{m}$ 
 in the cases of small mixing angle (SMA) 
 and large mixing angle (LMA) MSW solutions \cite{MSW}. 

\section{T Violation in Neutrino Oscillation
}
\label{sec:hamiltonian}

\hspace*{\parindent}
We review T violation in three-neutrino oscillations and 
state the strategy of this letter. 
In vacuum, flavor eigenstates $\nu_\alpha$($\alpha = e, \mu, \tau)$
are related to mass eigenstates $\nu_i$($i=1,2,3$), 
which have the mass eigenvalues $m_i$, by the unitary transformation, 
\begin{eqnarray}
  \label{eq:uai}
  \nu_{\alpha} = U_{\alpha i} \nu_i, 
\end{eqnarray}
where $U_{\alpha i}$ is the Maki-Nakagawa-Sakata matrix \cite{MNS}. 
The T violating part, 
$\Delta P_T
(\nu_{\alpha} \to \nu_{\beta}) \equiv
  P (\nu_{\alpha} \to \nu_{\beta}) 
- P (\nu_{\beta} \to \nu_{\alpha})
$, 
 in three generation after traveling a distance $L$ 
is calculated as 
\begin{equation}
  \label{eq:pt} %
  \Delta P_T
  =
 16 J
 \sin \frac{\Delta_{12}L}{2}
 \sin \frac{\Delta_{23}L}{2}
 \sin \frac{\Delta_{31}L}{2}, 
\end{equation}
where 
\begin{eqnarray}
  \label{eq:J} %
   J \!\!& \equiv &\!\! \mbox{Im}[U_{\alpha i}   U_{\beta i}^*
                            U_{\alpha j}^* U_{\beta j}]
.
\end{eqnarray}
In order to obtain the T violating part in matter,  
we only have to replace
$\Delta_{ij} \to (\Delta_m)_{ij}$, 
$
  U_{\alpha i} \to (U_m)_{\alpha i}$, 
hence, 
$
J \to J_{m}$.

We would like to study the case where 
large
$\Delta P_T$ 
is realized. 
In eq. (\ref{eq:pt}), 
$\Delta P_T$ 
is a product of 
 $J_{m}$  and  trigonometric functions. 
In the following calculation, 
 we focus on the matter effect of $J_{m}$
 which does not depend on $L$ 
 and determine the maxima of $J_{m}$. 

As seen in eq. (\ref{eq:J}), 
 $J$ consists of the product of 
$
U_{\alpha i}
$. %
It is complicated to calculate  $J_{m}$ directly 
from $(U_m)_{\alpha i}$, which diagonalizes the matter-modified Hamiltonian
$H_m$, although the numerical calculation has been performed
\cite{Xing}.
However, 
it is possible to calculate $J_{m}$ 
without direct calculation of 
$
(U_m)_{\alpha i}
$
from the relation 
\begin{eqnarray}
  \label{eq:harrison-scott} %
  (\Delta_m)_{12} (\Delta_m)_{23} (\Delta_m)_{31}J_{m}
  =
  \Delta_{12} \Delta_{23} \Delta_{31}J_{}, 
\end{eqnarray}
derived by Harrison and  Scott
\cite{harrison-scott}. 
Since the right hand-side of eq. (\ref{eq:harrison-scott}) 
is a constant which does not depend on the matter effect, 
$J_{m}$ is inversely proportional to a triple product of
$(\Delta_m)_{ij}$.
Therefore, we
 study the function of the matter potential $a$ such
as 
\begin{eqnarray}
  \label{eq:jmj}
  \label{eq:fa} %
   f(a) & \equiv & 
    [(\Delta_m)_{12} (\Delta_m)_{23} (\Delta_m)_{31}]^2, 
\end{eqnarray}
and determine the minima of $f(a)$.

\section{
Triple Product of Mass Square Differences
}
\label{sec:def-fa}

\hspace*{\parindent}
In this section, we study the matter effect on $f(a)$.
Harrison and Scott \cite{harrison-scott} suggest that $f(a)$ is
a quartic function of the matter potential $a$ 
and in principle its coefficients can be written by the parameters
$\Delta m^2_{ij}$ and $U_{\alpha i}$ in vacuum, 
although it is complicated in practice. 
We present the exact form of $f(a)$ 
 in relatively simple form 
 by introducing new parameters.
The coefficients of $f(a)$ is further simplified 
 under the approximation $|\Delta m^2_{12}| \ll |\Delta m^2_{13}|$.

First, let us note that $(\Delta_m)_{ij}$
 included in $f(a)$ are rewritten by 
 the eigenvalues $(\lambda_m)_{i}$
 of the matter-modified Hamiltonian 
 $H_m$  as $(\Delta_m)_{ij}=(\lambda_m)_{j}-(\lambda_m)_{i}$.
The eigenvalues $(\lambda_m)_{i}$ are the solutions of the equation for
 $t$, 
\begin{eqnarray}
  \label{eq:detH}
  \det(H_m-t)
=(\lambda_1-t)(\lambda_2-t)(\lambda_3-t)+a(t-\delta_2)(t-\delta_3)=0, 
\end{eqnarray}
where $\delta_{i}(i=2,3)$ and $\lambda_i(i=1,2,3)$ are the eigenvalues of
the $2\times2$ submatrix $H_{ij}(i,j=2,3)$ and 
$3\times3$ matrix $H$ 
 in vacuum. 
After a calculation, $f(a)$ is expressed as a quartic function of $a$: 
\begin{eqnarray}
  \label{eq:fa-eigenm} %
  f(a) & = & 
     [
          ((\lambda_m)_{2} - (\lambda_m)_{1})
          ((\lambda_m)_{3} - (\lambda_m)_{2})
          ((\lambda_m)_{1} - (\lambda_m)_{3})
     ]^2
\\ & = & 
  f_4 a^4 + f_3 a^3 + f_2 a^2 + f_1 a + f_0, 
  \label{eq:fa-eigen} %
\end{eqnarray}
where the coefficients $f_i$ $(i=0,\ldots,4)$ 
are presented by $\lambda_{i}$ and $\delta_{i}$
in the following.

The coefficients
 $f_4$ and
 $f_0$ are
\begin{eqnarray}
  \label{eq:fc4}
f_4 \!\!& = &\!\! 
        ( \delta_2 - \delta_3)^2, 
        \\
  \label{eq:fc0}
 f_0 \!\!& = &\!\! \mbox{ }
          \{
            (\lambda_{2} - \lambda_{1})
            (\lambda_{3} - \lambda_{2})
            (\lambda_{1} - \lambda_{3})
          \}^2. 
\end{eqnarray}
By definition (\ref{eq:fa}), $f(a)$ is semi-positive definite, 
hence, $f_4, f_0 \ge 0$ must be satisfied taking account of the limit
$a \to \infty$ and $a \to 0$.
The relations (\ref{eq:fc4}) and (\ref{eq:fc0}) are consistent with
these conditions.

The other coefficients are
\begin{eqnarray}
  \label{eq:fc3}
 f_3 \!\!& = &\!\! \mbox{ }
         2[
           (\delta_2 - \lambda_1)
           (\delta_2 - \lambda_2)
           (\delta_2 - \lambda_3)
          +
           (\delta_3 - \lambda_1)
           (\delta_3 - \lambda_2)
           (\delta_3 - \lambda_3)
         ]
\nonumber 
         \\ & & \hspace{2em}
         -
          2 (\delta_2 - \delta_3)^2
          [
             (\delta_2 - \lambda_1)
           + (\delta_3 - \lambda_1)
                + (\mbox{cyclic of }\lambda_i)
          ], 
\\
  \label{eq:fc2}
 f_2 \!\!& = &\!\! \mbox{ }
            [
              (\delta_2 - \lambda_1)
              (\delta_3 - \lambda_2)
                + (\mbox{cyclic of }\lambda_i)
            ]^2 
\nonumber 
         \\ & & \hspace{1em}
           - 
           6 [
              (\delta_2 - \lambda_1)(\delta_2 - \lambda_2)
               \{
               (\delta_3 - \lambda_1) +
               (\delta_3 - \lambda_2)
               \}
               (\delta_3 - \lambda_3)
            + (\mbox{cyclic of }\lambda_i)
             ]
 ,            \qquad
\\
  \label{eq:fc1}
 f_1 \!\!& = &\!\! \mbox{ }
            4 [
              (\delta_2 - \lambda_2)
              (\delta_3 - \lambda_2)
              (\lambda_{1} - \lambda_{3})^2
              (\lambda_{2} - \lambda_{1})
\label{eq:a-2-l-l-2} %
              + (\mbox{cyclic of $\lambda_{i}$})
              ]
           \qquad
\nonumber 
         \\ & & \hspace{2em}
         + 2 (\lambda_{2} - \lambda_{1})
             (\lambda_{3} - \lambda_{2})
             (\lambda_{1} - \lambda_{3})
             [(\delta_2 - \lambda_1)(\delta_3 - \lambda_2)
              + (\mbox{cyclic of $\lambda_{i}$})
             ], 
\label{eq:a-2-l-l-l} %
\end{eqnarray}
which are relatively simple compared with the case
where we don't introduce new parameters $\delta_{i}$. 
 In section $\ref{sec:experiment}$, 
we present the figures using these coefficients. 

Next, let us show that these coefficients are further simplified 
 under the approximation $|\Delta_{12}| \ll |\Delta_{13}|$. 
As 
$\delta_{i}$ are 
the eigenvalues 
 of submatrix 
 \begin{eqnarray}
   \label{eq:mu-tau-block} %
  \left(\begin{array}{cc} 
 H_{22} & H_{23}  \\
 H_{32} & H_{33}  \\
        \end{array}\right)
 =
   \lambda_{1}   
   {\bf 1}
  +
  \Delta_{13} 
  \left(\begin{array}{cccc} 
   |U_{\mu 3}|^2 &
   U_{\mu 3}U_{\tau 3}^*
   \\
   U_{\tau 3}U_{\mu 3}^* &
   |U_{\tau 3}|^2   
        \end{array}\right)
  +
  \Delta_{12}  
  \left(\begin{array}{cccc} 
   |U_{\mu 2}|^2  &
   U_{\mu 2}U_{\tau 2}^*
   \\
   U_{\tau 2}U_{\mu 2}^* &
   |U_{\tau 2}|^2   
        \end{array}\right)
, 
 \end{eqnarray}
 where ${\bf 1}$ is the unit matrix, 
they are approximated by
\begin{eqnarray}
\label{eq:delta-e} %
 \delta_{2} =
   \lambda_{1}
   +
          \frac{
|U_{e1}|^2
 \Delta_{12}
          }{
                1- |U_{e3}|^2
          }
,
\quad
 \delta_{3} = 
   \lambda_{1}
   +
                (1- |U_{e3}|^2) \Delta_{13}
+
          \frac{
|U_{e2}|^2|U_{e3}|^2
 \Delta_{12}
          }{
                1- |U_{e3}|^2
          }
, 
\end{eqnarray}
 up to the first order of $\Delta_{12}$
using the unitarity condition.
Substituting eq. (\ref{eq:delta-e})
 for $\delta_i$
 in 
eqs. (\ref{eq:fc4})$\sim$(\ref{eq:fc1})
 and taking the standard
parameterization,  
$U_{e1}=c_{12} c_{13}$, 
$U_{e2}=s_{12} c_{13}$, 
$U_{e3}=s_{13} e^{-i\delta}$, 
 where 
$c_{ij} = \cos \theta_{ij}$
 and 
$s_{ij} = \sin \theta_{ij}$, 
 the coefficients are calculated as 
\begin{eqnarray}
  \label{eq:fc-leading-PDG}
f_4 \!\!& \simeq &\!\! c^4_{13}(\Delta_{13})^2
 , \quad %
   f_3 \simeq
  -
  2c^4_{13} \cos 2 \theta_{13}  (\Delta_{13})^3
 , \quad %
  f_2 \simeq
 c^4_{13}(\Delta_{13})^4
 , \quad %
 \nonumber \\
 & & \quad
   f_1 \simeq
  -
 2 c^2_{13} \cos 2\theta_{12} 
\Delta_{12}(\Delta_{13})^4
 , \quad %
   f_0 \simeq
(\Delta_{12})^2(\Delta_{13})^4
, 
\end{eqnarray}
at the leading order. 
Note that 
the order of 
 $\Delta_{12}$ 
 for $f_{i}$ is important when we determine the minima of $f(a)$. 
$f_1$ is the first order of $\Delta_{12}$ and
$f_2, f_3, f_4$ are the zeroth order. 
Its difference determines the magnitude of 
 $a$
 for each minima. 

\section{Matter Enhancement of the Jarlskog Factor}
\label{sec:max}

\hspace*{\parindent}
In this section, 
we calculate the minima of 
 $f(a)$ 
 using the coefficients 
 (\ref{eq:fc-leading-PDG})
 in order to determine the maxima of $J_{m}$. 
First, we show that the number of minima depends
 on the 
 magnitude of 
$\sin^2 2 \theta_{13}$, 
 and that there are two minima 
 taking account of the constraint on $\sin^2 2 \theta_{13}$ 
 from the CHOOZ experiment. 
Second, 
we estimate the maximal values of $J_{m}$ and the energies of the
neutrino at maxima 
 in the cases of the SMA and LMA MSW solutions. 

Let us start with  
 differentiating $f(a)$ in terms of $a$: 
\begin{equation}
  \label{eq:fc-d-leading-PDG}
  f(a)'
 = 
 4 f_4 a^3
+ 3 f_3 a^2
+ 2 f_2 a
+   f_1
= 0
. 
\end{equation}
Since only $f_1$ is $O(\Delta_{12})$ in $f_i(i=1,2,3,4)$ 
from eq. (\ref{eq:fc-leading-PDG}), 
in the limit of $\Delta_{12} \to 0$, 
eq. (\ref{eq:fc-d-leading-PDG}) reduces to 
\begin{equation}
  \label{eq:fc-d-0}
 a (
 4 f_4 a^2
+ 3 f_3 a
+ 2 f_2
)
=0. 
\end{equation}
Hence, there exists a solution at $a=0$ in this limit. 
This means that a solution at $a=O(\Delta_{12})$ exists for
 $\Delta_{12} \ne 0$. 

On the other hand, 
whether another minimum exists or not
 is determined by 
  the discriminant $D$ of the quadratic equation
 in the parenthesis of eq. (\ref{eq:fc-d-0}), 
\begin{equation}
  \label{eq:fc-d-leading-PDG-0}
D=
 9 f_3^2
- 32 f_4 f_2
  = 4 c_{13}^8 (\Delta_{13})^{12}(1-9\sin^2 2 \theta_{13})
  .
\end{equation}
If 
 $
 \sin^2 2 \theta_{13} > 1/9
 $, 
there exists only one minimum at $a=O(\Delta_{12})$ 
 as Fig.~1 (a). 
If 
 $
 \sin^2 2 \theta_{13} < 1/9
 $, 
 then
there exists another minimum at $a=O(\Delta_{13})$
 as Fig.~1 (b). 
The restriction of the CHOOZ experiment, 
 $
  \sin^2 2 \theta_{13} \le 0.10
 $
\cite{CHOOZ}, 
is included in the case of Fig.~1 (b).  

\vspace*{1em}
  \begin{center}
 \begin{minipage}{\textwidth}
  \begin{center}
 \begin{minipage}{\pagewidthhalf}
  \begin{center}
  \includegraphics[bb=75 635 232 760]{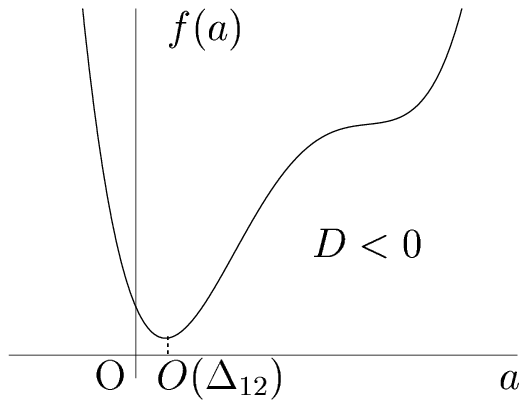}
  \end{center}
\vspace{-2em}
\begin{center}
{\small 
  (a) $D<0$ $(\sin^2 2 \theta_{13}>1/9)$
}
\end{center}
 \end{minipage}
 \begin{minipage}{\pagewidthhalf}
  \begin{center}
  \includegraphics[bb=75 635 232 760]{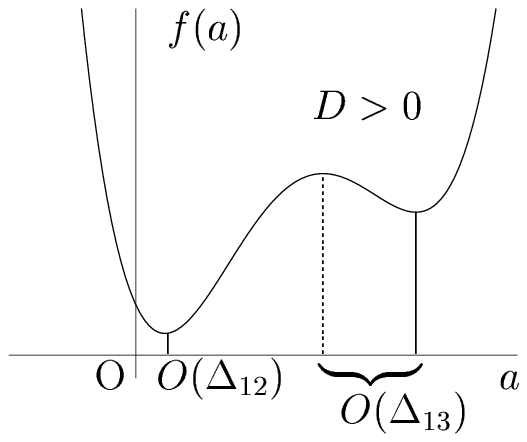}
  \end{center}
\vspace{-2em}
\begin{center}
{\small 
  (b) $D>0$ $(\sin^2 2 \theta_{13}<1/9)$
}
\end{center}
 \end{minipage}
    \end{center}
 \end{minipage}

\vspace{1em}
  \begin{minipage}[h]{\textwidth}
{\small 
Fig.~1.  
$f(a)$ has two(one) local minima for $D >0$ $(D < 0)$.
The CHOOZ experiment favors $D>0$ and the figure (b).
}
  \end{minipage}
   \end{center}

  \begin{flushleft}
{\bf (I) The maximal value of $J_m$ at $a = O(\Delta_{12})$}
  \end{flushleft}
\label{d12}

The solution of eq. (\ref{eq:fc-d-leading-PDG}) 
 at $a=O(\Delta_{12})$ is 
\begin{equation}
  \label{eq:a0-0}
  a =
   \frac{\cos 2 \theta_{12}}{\cos^2 \theta_{13}} \Delta_{12}.
\end{equation}
The minimal value
 is
\begin{equation}
  \label{eq:a0-0-v}
  f(a)
 =
       \sin^2 2 \theta_{12}
       (\Delta_{12})^2 (\Delta_{13})^4, 
\end{equation}
and thus, from eq. (\ref{eq:harrison-scott}), 
the maximum of the ratio is given by 
\begin{equation}
  \label{eq:a0-0-jj-PDG}
   \frac{J_{m}}{J}
 = \frac{1}{\sin 2 \theta_{12}}, 
\end{equation}
which is consistent with other works \cite{Xing,Yasuda}.
This means that $J_{m}$ is largely enhanced 
 in the case of the SMA MSW solution.

We estimate $J_{m}/J$ in 
 two MSW solutions: 
\begin{eqnarray}
  \label{eq:a0-Jm/J}
             \frac{J_{m}}{J}
 = 
\left\{
  \begin{array}{rl}
\displaystyle{
 12
}
,
 &
 \mbox{ for SMA MSW,}
\\
\displaystyle{
     1.1
,
}
 &
 \mbox{ for LMA MSW, }
  \end{array}
\right.
\end{eqnarray}
 where we use 
 $\sin^2 2 \theta_{12}=
7.2\times10^{-3}$$\mbox{ (SMA MSW), }$
$0.79$$\mbox{ (LMA MSW) }
$
\cite{BKA}.

The neutrino energy corresponding to the maximum of 
 $J_{m}/J$, from eq. (\ref{eq:a0-0}), is 
\begin{equation}
  \label{eq:a0-0-E}
  E =
\frac{\cos 2 \theta_{12}\Delta m^2_{12}}{2 \sqrt2 G_F N_e
   c_{13}^2}, 
\end{equation}
where $N_e$ is 
 the electron number density: 
 $N_e=8.2 \times 10^{23} $cm${}^{-3}$ in the earth's crust.
Substituting the experimental data, it is obtained as 
\begin{eqnarray}
  \label{eq:a0-0-E-v}
           E
 = 
\left\{
  \begin{array}{rl}
\displaystyle{
 25 \mbox{ MeV}
}
,
 &
 \mbox{ for SMA MSW,}
\\
\displaystyle{
     62 \mbox{ MeV}
,
}
 &
 \mbox{ for LMA MSW,}
  \end{array}
\right.
\end{eqnarray}
where
we use 
$\Delta m^2_{12}=
5.0\times10^{-6}$$\mbox{ (SMA MSW), }$
$2.7\times10^{-5}$$\mbox{ (LMA MSW), }
$
 and 
$\sin^2 2 \theta_{13}
 = 0.10$(the upper limit of the CHOOZ experiment). 

  \begin{flushleft}
{\bf (II) The maximal value of $J_m$ at $a = O(\Delta_{13})$}
  \end{flushleft}
\label{d13}

The other solutions of
eq. (\ref{eq:fc-d-leading-PDG})
at $a = O(\Delta_{13})$
 are
 \begin{equation}
   \label{eq:a0-n-0-PDG}
   a = \frac14 \left( 3 \cos 2 \theta_{13} 
         \pm \sqrt{1 - 9 \sin^2 2 \theta_{13}}
         \right) \Delta_{13}, 
 \end{equation}
where the sign 
  $+$ for $\cos 2 \theta_{13} \ge 0$
  and the sign 
  $-$ for $\cos 2 \theta_{13} \le 0$. 
The minimal value
 is
\begin{equation}
  \label{eq:fa-a0-n-0-v-PDG}
  f(a)
 =
     \frac{
      c^4_{13} (\Delta_{13})^6
     }{32}
     \left[
        4-3(1-3\sin^2 2 \theta_{13} )^2
        - \cos 2 \theta_{13} (1-9\sin^2 2 \theta_{13})^{\frac32}
     \right]
\end{equation}
 and 
 the maximum of the ratio 
 is given by 
\begin{equation}
  \label{eq:a0-n-0-jj-PDG}
  \frac{J_{m}}{J}
 =
     \left|
     \frac{
      \Delta_{12}
     }{\Delta_{13}}
     \right|
     \frac{
      1
     }{c^2_{13}}
      \frac{4\sqrt2}{
     \sqrt{
        4-3(1-3\sin^2 2 \theta_{13} )^2
        - \cos 2 \theta_{13} (1-9\sin^2 2 \theta_{13})^{\frac32}
     }
    }. 
\end{equation}
Because of 
the suppression factor 
 $|\Delta_{12}/\Delta_{13}|$, 
the enhancement of $J_{m}$ is small compared with the case (I).

Furthermore, we can obtain more simple forms
for eq. (\ref{eq:a0-n-0-PDG})
 and eq. (\ref{eq:a0-n-0-jj-PDG})
under the approximation
 $9 \sin^2 2 \theta_{13} \ll 1$,
although
 this approximation
 is not 
 justified near the
 upper limit,  
 $\sin^2 2 \theta_{13} \simeq 0.10$,  
 of the CHOOZ experiment.
In this case, 
the value of $a$ for the maximum of $J_{m}$ is
\begin{equation}
  \label{eq:a0-n-0-PDG-n}
  a
  =
    \left( 1 - \frac32 \sin^2 2 \theta_{13}
    \right) \Delta_{13}
\end{equation}
 and the ratio is 
\begin{equation}
  \label{eq:a0-n-0-jj-max-PDG-n}
            \frac{J_{m}}{J}
 = 
       \left|
       \frac{\Delta_{12}}{\Delta_{13}}
       \right|
       \frac{1}{\sin 2 \theta_{13}}.
\end{equation}
It is understood from
 this result 
 that the enhancement of $J_m$ is not always realized 
because of the suppression factor 
$|{\Delta_{12}}/{\Delta_{13}}|$. 
However, we still have an enhancement for small $\sin 2 \theta_{13}$. 
For example, at 
$\sin^2 2 \theta_{13}=4.0\times10^{-6}$ 
 which corresponds to 
$\sin \theta_{13}=1.0\times10^{-3}$ 
 and 
 which is much smaller than the present upper limit, 
the maximum of the ratio 
 is given by
\begin{equation}
  \label{eq:a0-n-0-Jm/J-n}
             \frac{J_{m}}{J}
 = 
\left\{
  \begin{array}{rl}
\displaystyle{
 0.78
}
,
 &
 \mbox{ for SMA MSW,}
\\
\displaystyle{
 4.2
,
}
 &
 \mbox{ for LMA MSW,}
  \end{array}
\right.
\end{equation}
where we use the experimental values
 for 
 $
\Delta m^2_{23}
=3.2
\times 10^{-3}\mbox{eV}^2$. 
Thus, 
$J_{m}$
 for the LMA MSW solution has an enhancement
which is several times as large as $J$
in this example. 

The neutrino energy corresponding to 
 the maximum of $J_{m}$ is calculated from eq. (\ref{eq:a0-n-0-PDG-n}), 
\begin{equation}
  \label{eq:a0-n-0-n-E}
E= {\left(1-\frac32\sin^2 2 \theta_{13}\right)}
   \frac{\Delta m^2_{13}}{2 \sqrt2 G_F N_e}. 
\end{equation}
Substituting the same experimental data as before 
we obtain
\begin{equation}
  \label{eq:a0-n-0-E-n}
           E
 = 
 16 \mbox{ GeV}
.
\end{equation}
We summarize the above two maxima of $J_{m}$
 in Table \ref{tab:jj-max}. 
\begin{table}[htbp]
\newcommand{\lw}[1]{\smash{\lower2.0ex\hbox{#1}}}
\begin{center}
 \begin{tabular}{|c|c|c|c|c|} \hline 
   $a$ &
    \multicolumn{2}{|c|}{$J_{m}/J$ and $E$} &
   SMA MSW &
   LMA MSW 
 \\ 
  \hline
   \lw{$O(\Delta_{12})$} & 
 $J_{m}/J$ &
 $1 / \sin 2 \theta_{12}$ &
 $12$ &
 $1.1$ 
\\ \cline{2-5}
  & $E$ &
$\cos 2 \theta_{12}\Delta m^2_{12}/(2 \sqrt2 G_F N_e c_{13}^2)$ &
  $24$MeV &
  $60$MeV 
\\
  \hline
   \lw{$O(\Delta_{13})$} & 
 $J_{m}/J$ &
 $\Delta m^2_{12} /(\Delta m^2_{13} \sin 2 \theta_{13})$ &
 $0.78$ &
 $4.2$ 
\\ \cline{2-5}
  & $E$ & 
  $[1-(3/2)\sin^2 2 \theta_{13}]\Delta m^2_{13}/(2 \sqrt2 G_F N_e)$ & 
  $16$GeV & 
  $16$GeV 
 \\ \hline
 \end{tabular}
\end{center}
\caption{
 The maxima of $J_{m}/J$ and the neutrino energies $E$. 
 Input parameters for vacuum are the same as in the text, 
 and
 $\sin^2 2 \theta_{13}=4.0 \times 10^{-6}$
  is chosen. 
}
\label{tab:jj-max}
\end{table}

\section{Numerical Estimation of the Ratio $J_m/J$}
\label{sec:experiment}

\hspace*{\parindent}
In this section, 
we numerically study the dependence of the ratio $J_{m}/J$ 
 on the neutrino energy $E$ 
 using 
 eqs. (\ref{eq:fa-eigen})$\sim$(\ref{eq:mu-tau-block}). 
First, 
we illustrate the magnitude of maximum for $J_m$
 taking account of 
 $\sin^2 2\theta_{12}$ and $\Delta m^2_{12}$ 
 given by two MSW solutions 
and 
  the constraint on $\sin^2 2\theta_{13}$
 from the CHOOZ experiment. 
Second, 
we study the effect of 
 the signs of $\Delta_{12}$ and/or $\Delta_{13}$. 
In the following calculation, 
input parameters are the same as in the previous section 
and 
we 
 restrict to the range 
$0<\theta_{ij} \le \pi/4$ for simplicity.

Let us show 
 the energy $E$ dependence of $J_{m}/J$ 
 in the cases of
 the SMA and LMA MSW solutions 
 with 
 $\sin \theta_{13}=|U_{e3}|=0.16$
 and 
 $|U_{e3}|=1.0 \times 10^{-3}$ 
 in Fig.~2 (a)$\sim$(d)
\footnote{The energy dependence of $J_{m}/J$ for a LMA MSW solution with
$|U_{e3}|$=0.090 (corresponding to Fig. 2(b)) 
is shown in ref.\cite{Xing}.}. 
\vspace*{-0.3em}
\begin{center}
 \begin{minipage}{\textwidth}
         \includegraphics[width=\pagewidthhalf-0.2cm,bb=50 50 230 176]{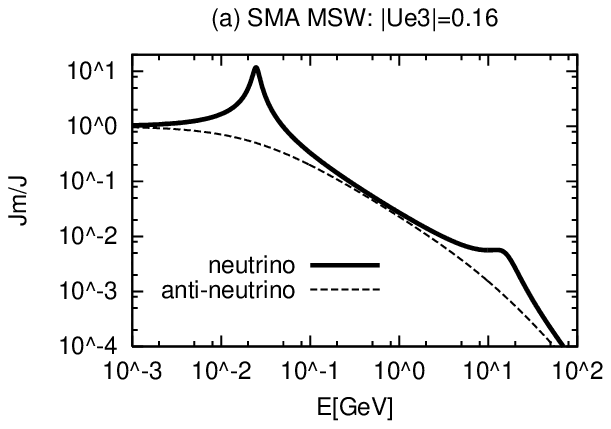}
         \includegraphics[width=\pagewidthhalf-0.2cm,bb=50 50 230 176]{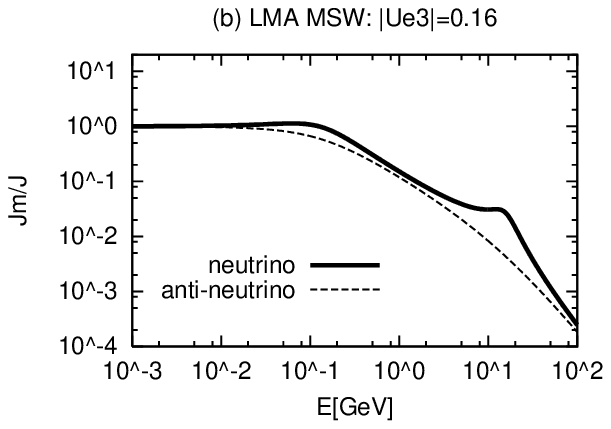}
  \end{minipage}
\end{center}
\begin{center}
  \begin{minipage}{\textwidth}
         \includegraphics[width=\pagewidthhalf-0.2cm,bb=50 50 230 176]{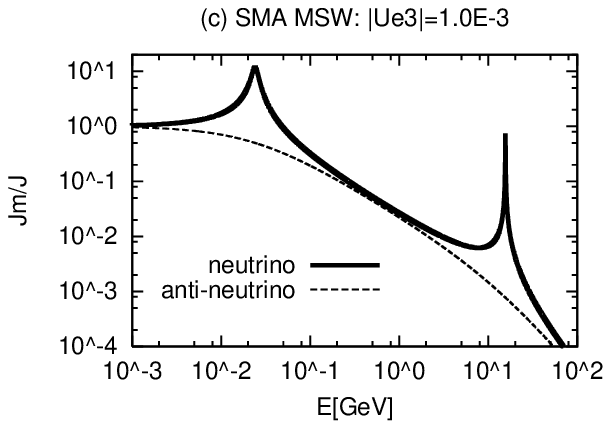}
         \includegraphics[width=\pagewidthhalf-0.2cm,bb=50 50 230 176]{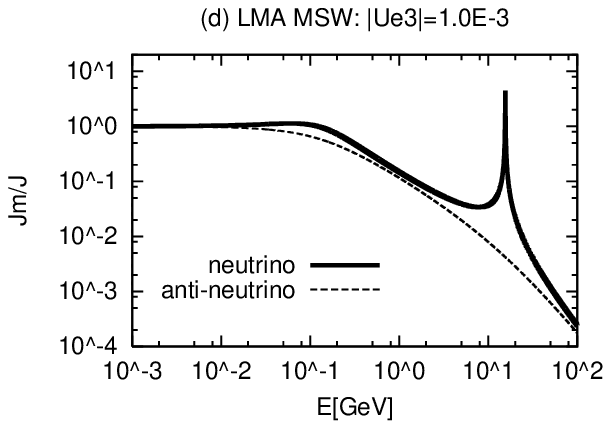}
  \end{minipage}
\\ \vspace*{0.3cm}
  \begin{minipage}{\textwidth}
{\small 
Fig.~2. 
 The neutrino energy $E$ dependence of $J_{m}/J$ 
 in the cases of the SMA and LMA MSW solutions with
 $|U_{e3}|=0.16$
and $|U_{e3}|=1.0\times10^{-3}$. 
The symbol $+$ denotes the maxima determined in the previous section.
}
  \end{minipage}
\end{center}
Comparing Fig.~2 (a) with (b) (or Fig.~2 (c) with (d)), 
 we conclude that the SMA MSW solution has larger enhancement than
 the LMA MSW solution has
 for 
the maximum of $J_{m}/J$ at $E=O(10$MeV$)$. 
The enhancement, $J_{m}/J >1$, occurs in the wide energy region
 around this maximum
 and the values calculated numerically almost coincide with
 the results (\ref{eq:a0-Jm/J}) and (\ref{eq:a0-0-E-v}) obtained approximately.  
Next, 
comparing Fig.~2 (a) and (c) (or Fig.~2 (b) and (d)),
we conclude that if $\sin \theta_{13}$ is small enough, 
 the enhancement for the maximum of $J_{m}/J$ at $E=O(10$GeV$)$ is
 possible 
although the energy region is small. 

Next, 
we study the cases where 
 $\Delta m^2_{12}$ and/or
 $\Delta m^2_{13}$ is negative. 
 Since
 we have implicitly assumed that 
 both $\Delta m^2_{12}$ and $\Delta m^2_{13}$ 
 are 
 positive
 until now, 
 two maxima appear in ``neutrino'' oscillation. 
 However, exactly speaking, 
 whether the maxima appear in ``neutrino'' oscillation or
 ``anti-neutrino'' oscillation depends on the signs of 
 $\Delta m^2_{12}$ and $\Delta m^2_{13}$. 
In order to examine such cases, 
we numerically calculate $J_m/J$ in the cases where
 $\Delta m^2_{12}$ and
  $\Delta m^2_{13}$ respectively, are
 positive and/or negative, 
 and show the results for the SMA MSW solution as an example in Fig.~3. 

\vspace*{-0.1em}
\begin{center}
 \begin{minipage}{\textwidth}
         \includegraphics[width=\pagewidthhalf-0.2cm,bb=50 50 230 176]{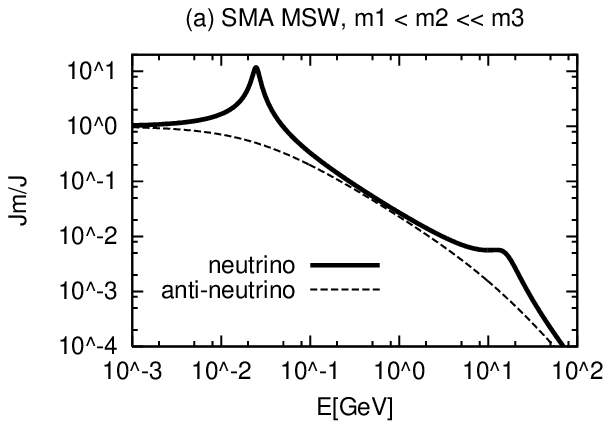}
         \includegraphics[width=\pagewidthhalf-0.2cm,bb=50 50 230 176]{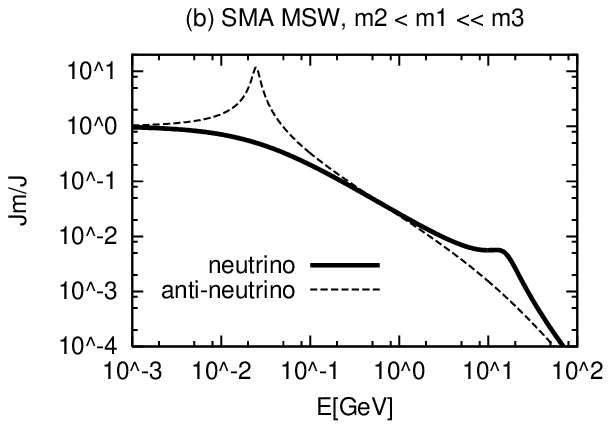}
  \end{minipage}
\end{center}
\begin{center}
  \begin{minipage}{\textwidth}
         \includegraphics[width=\pagewidthhalf-0.2cm,bb=50 50 230 176]{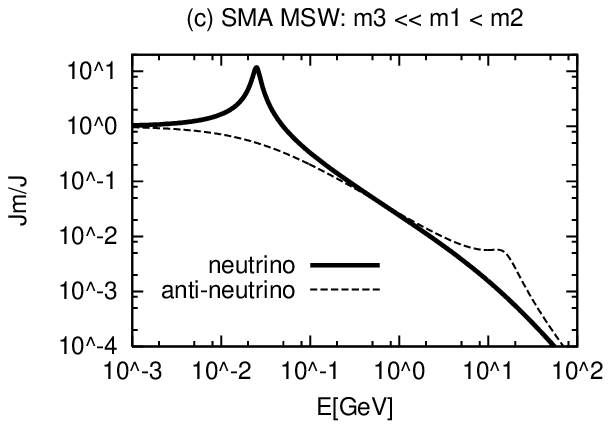}
         \includegraphics[width=\pagewidthhalf-0.2cm,bb=50 50 230 176]{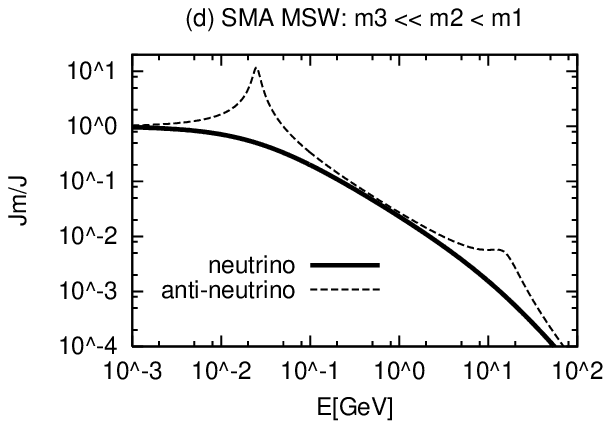}
  \end{minipage}
\\ \vspace*{0.3cm}
  \begin{minipage}{\textwidth}
{\small 
Fig.~3. 
The dependence of $J_{m}$ on the sign of $\Delta m^2_{ij}$.
(a) and (c) are for $\Delta m^2_{12}>0$
, 
(b) and (d) for $\Delta m^2_{12}<0$. 
(a) and (b) are for $\Delta m^2_{13}>0$
, 
(c) and (d) for $\Delta m^2_{13}<0$. 
The other conditions are the same as in Fig.~2(a). 
}  \end{minipage}
\end{center}

  Comparing Fig.~3 (a) with (b)
  (or Fig.~3 (c) with (d)), 
 we conclude that the appearance of the maximum for $J_{m}/J$
 at $E=O(10$MeV$)$ depends 
 on the sign of $\Delta m^2_{12}$. 
 Although
 the maximum 
 appears in ``neutrino'' oscillation
 in the case of $\Delta m^2_{12} >0$, 
 it appears in ``anti-neutrino'' oscillation
 in the case of $\Delta m^2_{12} <0$.
  Comparing Fig.~3 (a) with (c) (or Fig.~3 (b) with (d)), 
 we conclude that the appearance of the maximum for $J_{m}/J$
 at $E=O(10$GeV$)$ depends 
 on the sign of $\Delta m^2_{13}$. 
 The maximum 
 appears in ``neutrino'' oscillation in the case of $\Delta m^2_{13} >0$. 
 On the other hand, 
 it appears in ``anti-neutrino'' oscillation
 in the case of $\Delta m^2_{13} <0$.

These differences 
 originate from the fact that 
the matter potential $a$ for the maxima of $J_{m}$
(see eqs. (\ref{eq:a0-0}) and (\ref{eq:a0-n-0-PDG})) 
 is proportional to 
 $\Delta m^2_{ij}$. 
If $\Delta m^2_{ij}$ is negative, 
then $a$ is also negative and
 $J_{m}$ has the maximum not in ``neutrino'' oscillation
but in ``anti-neutrino'' oscillation. 
This is because 
 the matter-modified Hamiltonian for anti-neutrino
 is obtained 
 by replacing $a \to - a$.

   \section{Summary and Discussions}

\hspace*{\parindent}
In this letter, 
we have studied matter modified Jarlskog factor $J_{m}$
 which appears in T violation for the lepton sector
 in neutrino oscillation. 
It was shown \cite{harrison-scott} that 
the inverse of $J_{m}$ is
proportional to the square root of a quartic
 polynomial
 of matter potential $a$
from the relation
$
 (\Delta_m)_{12}
 (\Delta_m)_{23}
 (\Delta_m)_{31}
J_{m}
  =
  \Delta_{12} \Delta_{23} \Delta_{31}J_{}
$. 
We have presented 
 the exact form of this polynomial
 with parameters in vacuum 
 and have reconsidered the matter enhancement of $J_{m}$
under the approximation $|\Delta m^2_{12}|\ll|\Delta m^2_{13}|$.

We show that $J_{m}$ has 
 (i)
 one maximum at $a = O(\Delta_{12})$ 
 in the case of $\sin^2 2 \theta_{13} \ge 1/9$ 
 and 
 (ii)
 two maxima at $a = O(\Delta_{12})$ 
and $a = O(\Delta_{13})$ in the case of $\sin^2 2 \theta_{13} < 1/9$. 
 Considering the constraint on $\sin^2 2\theta_{13}$ 
 from the CHOOZ experiment, 
 we conclude that the case (ii) is realized. 

One maximum of $J_m$ at $a=
(\cos 2 \theta_{12}/\cos^2 \theta_{13})\Delta_{12}$ 
 is given by 
$J/\sin 2 \theta_{12}$. 
$J_{m}/J$
 is roughly estimated as 
 12
 for the SMA MSW, thus large enhancement is realized. 
The other maximum 
 at $a
=
(1-\frac32 \sin^2 2 \theta_{13})\Delta_{13}$ 
 is given by 
$|\Delta_{12}/\Delta_{13}| J / \sin 2 \theta_{13}$ 
 for $\sin^2 2 \theta_{13} \ll 1/9$. 
If $\theta_{13}$ is small enough, 
 the ratio $J_m/J$ is enhanced. 
We have roughly estimated $J_m/J$ 
 as 
 $4.2$ for the LMA MSW solution
 at $\sin^2 2 \theta_{13}=4.0 \times 10^{-6}$. 

In the case of $\sin^2 2 \theta_{13}=4.0 \times 10^{-6}$, 
our results agree with the results obtained by 
a different method \cite{Yasuda}. 
Our results are also applicable 
 around 
 $\sin^2 2 \theta_{13} \simeq 0.10$
which is the upper limit from the CHOOZ experiment.

\vspace{20pt}
\noindent
{\Large {\bf Acknowledgment}}

\noindent
The authors would like to thank Prof. A. I. Sanda for
useful 
 comments.
We would like to thank Dr. M. Harada for careful reading
of manuscript.


\begin{thebibliography}{99}

\bibitem{solar} 
Homestake Collaboration, B.~T.~Cleveland {\rm et al.}, 
Astrophys. J. {\rm 496}, 505 (1998); 
SAGE Collaboration, J.~N.~Abdurashitov {\rm et al.}, 
Phys. Rev. {\rm C60} (1999) 055801; 
GALLEX Collaboration,  W.~Hampel {\rm et al.}, 
Phys. Lett. {\rm B447} (1999) 127; 
Super-Kamiokande Collaboration, Y. Fukuda  {\rm et al.}, 
Phys. Rev. Lett. {\rm 82} (1999) 1810; 
 {\rm ibid.}, 82 (1999) 2430. 

\bibitem{atmo} 
IMB Collaboration, 
R.~Becker-Szendy {\rm et al.}, 
 Phys. Rev. {\rm D46} (1992) 3720; 
SOUDAN2 Collaboration, 
W.W.M.~Allison 
{\rm et al.}, Phys. Lett. {\rm B391} (1997) 491; 
{\rm ibid.}, Phys. Lett. {\rm B449} (1999) 137; 
SuperKamiokande Collaboration, Y. Fukuda {\rm et al.}, 
 Phys. Rev. Lett. {\rm 82} (1999) 2644; 
{\rm ibid.}, 
Phys. Lett. {\rm B467} (1999) 185. 

\bibitem{MNS}
Z.~Maki, M.~Nakagawa and S.~Sakata, 
 Prog. Theor. Phys. {\rm 28} (1962) 870. 

\bibitem{exp}
Y.~Oyama (K2K Collaboration), hep-ex/0004015; 
 OPERA Collaboration, K.~Kodama {\rm et al.}, 
 CERN/SPSC 99-20, SPSC/M635, LNGS-LOI 19/99 (1999); 
 ICARUS and NOE Collaborations, F.~Arneodo {\rm et al.}, 
 INFN/AE-99-17, CERN/SPSC 99-25, SPSC/P314 (1999); 
 MINOS Collaboration, P.~Adamson {\rm et al.}, 
 NuMI-L-337 (1998). 

 \bibitem{n-f-ex-review}
   S.~Geer, Phys. Rev. {\rm D57} (1998) 6989
; 
   A.~Cervera, {\rm et al.}, hep-ph/0002108
; 
A. De Rujula, M. B. Gavela, and P. Hernandez, 
 Nucl. Phys. {\rm B547} (1999) 21. 

\bibitem {CP-T}
J. Arafune and J. Sato, Phys. Rev. {\rm D55} (1997) 1653; 
J. Arafune, M. Koike and J. Sato, Phys. Rev. {\rm D56} (1997) 3093; 
Erratum {\rm ibid.}, {\rm D60} (1999) 119905; 
 J. Sato, Nucl. Phys. Proc. Suppl. 59 (1997) 262, 
M. Koike and J. Sato, 
 hep-ph/9911258; 
M. Tanimoto, Phys. Rev. {\rm D55} (1997) 322; 
{\rm ibid.}, Prog. Theor. Phys. {\rm 97} (1997) 901; 
H. Minakata and H. Nunokawa, Phys. Lett. {\rm B413} (1997) 369; 
{\rm ibid.}, 
 Phys. Rev. {\rm D57} (1998) 4403; 
S. M. Bilenky, C. Giunti and W. Grimus, 
 Phys. Rev. {\rm D58} (1998) 033001; 
K. Dick {\rm et al.}, Nucl. Phys. {\rm B562} (1999) 29. 

\bibitem {n-f-CP-matter}
M. Tanimoto, Phys. Lett. {\rm B462} (1999) 115; 
 A. Donini, M. B. Gavela, P. Hernandez, S. Rigolin, 
 Nucl.Phys. {\rm B574} (2000) 23; 
 V. Barger, S. Geer, K. Whisnant, 
 Phys. Rev. D61 (2000) 053004; 
M. Koike and J. Sato, Phys. Rev. {\rm D61} (2000) 073012; 
H.~Minakata, H.~Nunokawa, 
hep-ph/0004114. 

\bibitem{T}
 T. K. Kuo and J. Pantaleone, Phys. Lett. {\rm B198} (1987) 406; 
 P. I. Krastev and S. T. Petcov, Phys. Lett. {\rm B205} (1988) 84; 
 S. Toshev, Phys. Lett. {\rm B226} (1989) 335; 
 {\rm ibid.}, Mod. Phys. Lett. {\rm A6} (1991) 455. 

\bibitem{Jarlskog}
C. Jarlskog, Phys. Rev. Lett. {\rm 55} (1985) 1039. 

\bibitem{Xing} 
   Z-Z. Xing, 
hep-ph/0002246. 

 \bibitem{Yasuda} 
   O.~Yasuda, 
   Acta. Phys. Polon. {\rm B 30} (1999) 3089. 

  \bibitem{harrison-scott} 
   P.~F.~Harrison and W.~G.~Scott, 
   Phys. Lett. B476 (2000) 349. 

\bibitem {CHOOZ}
CHOOZ Collaboration, M. Apollonio {\rm et al.}, , 
Phys. Lett. {\rm B466} (1999) 415. 

\bibitem{MSW}
   S.~P.~Mikheev and A.~Yu.~Smirnov, 
   Sov. J. Nucl. Phys. {\rm 42}, 913 (1985);
   L.~Wolfenstein, Phys. Rev. {\rm D17} (1978) 2369.  


\bibitem{BKA} 
J. Bahcall, P. Krastev and A. Smirnov, 
 Phys. Lett. {\rm B477} (2000) 401. 

\end{thebibliography}
\end{document}